\documentstyle[prd,aps,12pt]{revtex}

\begin{document}

\draft
\title{Photoproduction of the Etas near Threshold}

\maketitle

Krusche {\it  et al}. have reported, in a recent Letter\cite{ref1},
precise measurements of differential and total cross sections on
the reaction $\gamma p\rightarrow p\eta$, from threshold to 790 MeV
of the lab photon energy. We draw here
some differing conclusions from those in the Letter, in interpreting
this data set.

The analysis in \cite{ref1} is based on its Eqs.(2) and (3). It assumes that
the E$_{0+}$ amplitude can be written
{\it entirely} in terms of the N$^*$(1535), which, in turn, can be expressed in
a simple Breit-Wigner (BW) form, Eq.(3). This contrasts with
our earlier study\cite{ref2} on the effective
Lagrangian approach (ELA) that included background contributions and
a more complicated
dependence of the resonance profile on the
cm energy W than a simple BW form \cite{ref1}. We compare these two
approaches here.

Our main point here is the following: although
the extraction\cite{ref1} of the helicity amplitude
A$_{1/2}$ for the N$^*$ (1535)
radiative decay to proton is model-dependent,
we can extract from the new data a parameter,
$\xi$\cite{ref2}, characteristic of the electrostrong property of
the baryon resonance N$^*$(1535) in a {\it model-independent}
manner.

To demonstrate this, we display, in Table I, the results of {\it eight
different fits}. First three rows  are fits to new differential
cross section data, last row to the reduced total one. The rows a (d) and b
represent two extreme cases in Table I of \cite{ref1},
involving the contribution of $N^*$ (1535) only, the row a (d)
corresponding to $W_R = 1549$ MeV, $\Gamma_R = 202$ MeV,
$b_{\eta} = 0.55$, the row b having these parameters
as 1539 MeV, 208 MeV and 0.35 respectively; the row c includes
background contributions\cite{ref2}
from the nucleon Born terms, vector meson
and the $N^*$ (1520) [D13] exchanges,
with $W_R$ and $\Gamma_R$ in the PDG recommended range\cite{ref3}.
In  each row, we give results for the ELA\cite{ref2}  and
BW\cite{ref1} fits.

The main conclusion from  our Table I is the following:
{\it while the extraction of A$_{1/2}$ is model-dependent,
the quantity $\xi$,
characteristic\cite{ref2} of the product of the strong and electromagnetic
amplitude for N$^*$(1535), remains quite stable.}
The  new data set\cite{ref1},  yields $\xi$\cite{ref2},
to be $2.20\pm 0.15$, in unit of $10^{-4}MeV^{-1}$.

Regarding the contributions of the other resonances, one should
go beyond the truncation  of Krusche {\it et al}.
in their Eq.(7) [note a sign error on their cos$\theta$ term].
We agree with their {\it qualitative} conclusion that the
presence of N$^*$(1520)[D13] is indicated, but its {\it quantitative} role
depends  on various background
contributions\cite{ref2}, ignored in [1]. The
inclusion of the N$^*$(1520)[D13] state, along with
the nucleon and vector meson exchange terms {\it does
improve} our fit
(the row c in Table I) to the {\it differential} cross section.
Futher exploration of this would require
polarization data, not yet
available.

We thank Prof. B. Krusche for
communicating the MAMI data to us. This work is supported by
the U. S. Department of Energy at RPI, and by the Natural
Sciences and Engineering Research Council of Canada at SAL.

Nimai C. Mukhopadhyay$^{a}$, J.-F. Zhang$^a$
and M. Benmerrouche$^{b}$\\
$^a$ Physics Department, Rensselaer Polytechnic Institute,
Troy, NY 12180-3590. \\
$^{b}$ Saskatchewan Accelerator Laboratory, University of Saskatchewan,
Saskatoon, SK S7N 0W0.


\newpage

\begin{table}
\caption{Fitted $A_{1/2}$
 and inferred $\xi$ for N$^*$(1535)
obtained from the new MAMI data\protect\cite{ref1}, along with
$\chi^2$ per degree of freedom, for various analyses explained
in the text. First three result columns are for ELA, while
the last three are for BW. First three rows are fits to the
differential cross sections, last row to the reduced total ones. }

\begin{center}
\begin{tabular}{|lrrrrrr|}
&$A_{1/2}$&$\chi^2$ & $\xi$ & $A_{1/2}$ & $\chi^2$ &$ \xi$ \\
a& $113$& 2.8& 2.2 &$111$& 2.2 & 2.2  \\
b&$144$& 3.0 & 2.3 & $142 $& 2.2 &2.3 \\
c&$98$& 1.4& 2.2&$100$& 1.5& 2.3 \\
d&$117$ & 3.8& 2.3 & $110$ & 0.9 & 2.2 \\
\end{tabular}
\end{center}
\end{table}

\end{document}